# A Review of Inferring Latent Attributes from Twitter

**Surabhi Singh Ludu**

Global Institute of Technology, Jaipur, Rajasthan, India
Affiliated to Rajasthan technical University, Kota, Rajasthan, India
*surabhisinghludu@gmail.com*

**Abstract:** *This paper reviews literature from 2011 to 2013 on how Latent attributes like gender, political leaning etc. can be inferred from a person's twitter and neighborhood data. Prediction of demographic data can bring value to businesses, can prove instrumental in legal investigation. Moreover, political leanings and ethnicity can be inferred from the wide variety of user data available on-line. The motive of this review is to understand how large datasets can be made from available twitter data. The tweeting and re tweeting behavior of a user can be user to infer attributes like, gender, age etc. We'll also try to understand the applications of Machine learning and Artificial Intelligence in this task and how it can be improved for future prospects. We explore in this text how this field can be expanded in future and possible avenues for future research.*

## 1. Introduction

Twitter is a micro blogging platform that has become a significant part of modern living with people from all walks of life using this website to express themselves. This results in a lot of user generated data, and thus the need and demand for analysis of this data is increasing. Twitter does not store data like age, gender, political leanings, and ethnicity. Many people in the field of research share an interest in breakdown of demographic attributes of twitter users. Inferring all these attributes from user generated data can be applied in providing the user a better user experience by improving the quality of who to follow, posts to read. The quality of relevant material like topics of interests, hashtags pertaining to a particular community can be improved. This data analysis can also be useful for brand placement based on gender, age, and other social demographic attributes for more accurate online marketing. Quantifying political leanings can yield better results in exit polls, and help in formulating strategies for targeting a particular demographic.

Many approaches have been used to classify twitter data like determining gender based on user's self-reported name [5] which yields 20% increase in accuracy, using neighborhood data i.e. along with data provided by a user, the postings by a follower or a friend, to increase accuracy by using the concept of homophily to infer attributes like political affiliation, age and gender [3], using retweet information to glean political leaning [4]. Some works used explore celebrity following tendency to determine the gender of users [6], [8]. This study aims at summarizing the methodologies used thus far for the purpose of data analysis of twitter based informal text.

## 2. Methodology

Literature from 2011 to 2015 inclusively were searched using the search terms such as review, inference twitter, latent attributes, Gender on Twitter, political leaning twitter, from ACM Digital Library, Association for the Advancement of Artificial Intelligence (www.aaai.org), Association for Computational Linguistics (www.aclweb.org), Journal of Sociolinguistics, etc. A total of 11 selected articles were reviewed. Primary data source for this review include peer-reviewed journal articles, conference proceedings etc. The references provide a complete list of all the articles and journals reviewed for this project. The publications reviewed are organized into four dimensions that address:

(a) Potentials of twitter based inference
(b) Different approaches used for different data of varying assortativity
(c) Research findings from studies using quantitative and/or qualitative methodologies
(d) Issues and challenges in twitter based inference

This categorization provides a framework to understand how data sets were formulated, data cleaning was done and then the algorithms used to arrive at a conclusion for a particular attribute.

Outline of this review was grouped into two particular aspects. The first aspect is inferring methodology and approach. This review aimed to understand how the twitter data inference has been applied. Findings in this aspect are reported based on dataset development, data sampling and cleaning and dividing data into training and testing sets. In earlier stages, manual classification was done for inferring attributes but it was neither scalable nor feasible for larger data sets. Therefore, the need for automated Machine learning models based on an algorithm like Support vector machine arose. Afterwards, a lot of studies of how this can be accomplished for various attributes were done. But this still remains a field of interest as many attributes can be inferred from user provided informal text data for usage in a lot of applications. Second aspect of this review attempted to analyze all research evidences that how this review can be applied in practical scenarios and what future avenues are there in this field.

## 3. Usage of Inferring Latent Attributes From Twitter

This section analyzes findings from other studies regarding how the SR technology was applied in learning in the past sixteen years. Finding in this section are classified into the following categories:

*3.1* **Advertising**

Advertising is a field that thrives on a user's likes and interests. Most products are marketed towards a particular demographic, for example toys are marketed towards kids. Latent attributes like age and gender can be leveraged for ad placement to target a larger user base.

*3.2* **Election Prediction**

A lot of methods like surveys and exit polls have been used traditionally for election predictions with varying degree of accuracy. In recent times people have taken to twitter to openly express their opinions on political matters which are rife with data that can be processed to infer an individual's political affiliations. So far, this has been applied to a two party system (e.g. USA), as a user can be clearly marked as belonging to either parties or neither.

*3.3* **Improved User Experience**

Most of the online platforms aim at improving user experience to keep user base growing. Data analysis findings from present user base can help improve user experience thus ensuring future growth for these platforms.

## 4. Literature Review

In [1] Burger, J.; Henderson, J.; and Zarrella, G.; describe an approach to treat gender classification as a binary classification problem. Data sampling for this project began as early as 2009 with a frequency of 400,000 tweets per day or 2% of twitter's daily volume at the time. Data was partitioned in three subsets, for training, development and test. Data fields used for the research were full name, screen name and description, although of all three only screen name is mandatory. A data field is a set of text strings as a user can either change or remove any the data any time. No language specific processing was performed, only tokenization done was at transition between numeric and alphanumeric characters. To speed up the process and compress data volume for then available machine learning toolkits, feature patterns are converted to integer code words. A lot of classifier types were used but balanced Winnow fetched most accurate results without much training time. Each of the fields were tried on the algorithm in isolation. Learning model was evaluated on three conditions: single tweet, all fields, and all tweets. Amazon Mechanical Turk was used to compare accuracy and results with human performance. The best classifier performed at 92% accuracy, and the classifier relying only on tweet texts performed at 76% accuracy.

In [2] Pennacchiotti, M., and Popescu, A. in their research on applying Machine learning approach to twitter classification leverage user profile, tweeting behavior, network and content from a user's feed to infer latent attributes. The framework they have developed can be used for three different tasks, i.e. Political orientation, ethnicity and business fan detection. Gradient Boosted Decision Trees - GBDT (Friedman 2001) has been used as a learning algorithm. Profile features such as user bio and location which user may or may not fill have been used for profile features and 30 regular patterns have been matched to generate ethnicity and gender of users under study. Tweeting behavior like average no of tweets per day, the number of replies etc. have also been used in the study. Linguistic content holding the main topics of interest has been explored, for e.g. younger populace tends to use 'lmao' and 'dude', while democrats use 'health-care' in their tweets. Hashtag is another linguistic feature exploited in the study. Friend accounts or who a user follows is also leveraged to find gender, ethnicity and political leanings. Results showed that identifying political affiliation was relatively easier than determining ethnicity.

In [3] Shadiev Zamal, F. A.; Liu, W.; and Ruths, D. describe in their paper the use of neighborhood data, i.e. followers and friends and people a user follows to determine latent features by applying the concept of homophily or the hypothesis that users with similar attributes cluster together in a social setting. It is claimed that neighborhood data alone can yield accuracy and precision at par or even better than user profile data. Different types of neighborhood definitions garner different results with varying success. Datasets of 400 users were collected with two labels, male and female, and it was manually confirmed that label assignments were correct. Ages were distinguished in groups (18+, 25+). User's feature vector had a set of N features and the same were computed for his neighborhood as well. Neighborhood feature was the average value of the features for all users in the neighborhood. Support vector machine algorithm was used as machine learning framework. To determine how neighborhood data yielded better results, four features were designed, user-only (all neighborhood features were omitted), neighbor-only (only the neighborhood features were used), averaged (each feature was the average over the neighborhood and the user feature values), and joined (the user features and neighborhood features were concatenated). While absolute accuracy improved from 3% to 5%, inference for age improved by 21% while political inference improved by 38%. It was also found that neighborhood selection requires attention while selection as not all neighbors fetch same results with same accuracy.

In [4] Wong, F.M.F.; Tan, C.W.; Sen, S.; Chiang M. apply a simple approach that does not need explicit knowledge of network topology. They talk about two challenges in this task, Quantification and scalability. They propose use of a method akin to link analysis for webpage ranking, i.e. to place a consistency condition between tweets and retweets and devise an inference technique on the same. Data used for this study is collected over seven months at the time of 2012 U.S. presidential elections. This can be applied not only for election prediction but also for providing a user balanced view on a controversial topic. Tweets are grouped by events, although other groupings can also be applied. Two average scores are computed, one based on tweets and other based on retweets. Based on tweet count events like presidential debate etc. were identified. Sentiment analysis of tweets was performed to extract sentiment of tweets. The results show that retweets can be classified to infer latent attributes.

In [5] Liu, W.; and Ruths, D. explore the use of first name of a user to determine its gender. They provide two different strategies for incorporating user's self-reported name as a classifier for gender. This study is based on the observation that most names have a set gender, and a few people are named outside of gender norms, and census data confirms this hypothesis. This method can be coupled with other methods of gender determination to provide increased accuracy in results. A gender labelled twitter dataset was built, and in a novel approach account profile picture was used to determine gender. Amazon Mechanical Turk was used for this purpose. It was found that including user's first name improves accuracy by 4% compared to baseline methods. They also released the dataset sampled for future use for community.

In [6] Ludu, P.S. explores the idea of inferring gender from analyzing what celebrity a user follows. This approach can be used for additional user classification. This work describes a set of popular neighborhood to classify gender of a user. Combination of both linguistic features and popular neighborhood proves to be perform better. Datasets used for research by Al Zamal et al. (2012)[3] was reused for this work, though some data cleaning was done manually. Popular neighborhood datasets were created with users having more than 10,000 followers marked as 'celebrities'. Other features used were Tweet behavior, tweet frequency, retweet frequency, celebrity following tendency etc. Popular neighborhood features can be extracted from Wikipedia or Freebase. On twitter users have more selectivity over who thy follow, so this data can be used to infer user features more accurately.

In [7] Bamman, D.; Eisenstein, J.; and Schnoebelen, T. study the relationship between linguistic style , social network and gender. While previous work marks this variable as either male or female, a different approach has been used. Different linguistic styles have been identified and clustered showing strong gender orientation.

In [8] Ludu, P.S. in his research builds on his previous work [6] and applies it to Indian populace using twitter. He uses the concept of class influencer, who are users who can influence a class so much so that they become a factor in their own right. Features like smiley type and count, night-time tweet frequency etc. This approach sees promising improvements in the accuracy of the results. Categorical words like 'money' , 'finance' etc. have also been exploited. For e.g. these words come up mostly in tweets by male users. Class identifiers are another discriminating factor, e.g. Twitter handle, 'MissMalini' is more popular among female Indian twitter users. Also age of users could be determined by the age and gender f the celebrities they followed. Similarly more BJP supports followed BJP leaders than AAP supporters. Updation of class identifiers is one feature that can be explored in future.

## 5. Conclusion and Future Work

It is clear that attribute inference from tweet based data can be classified based on can be accomplished using a lot of approaches. It is clear that there have been a lot of challenges in this field. Manual classification and data sampling was a time consuming and tedious process and thus it wasn't scalable. Another problem is that people often misrepresent themselves on many online social platforms. For example. Usage of a celebrity for a profile picture or using an alias as a name, using places such as, Neverland, wonderland etc. as locations. This can lead to false data interpretation which in turn can affect the results of a research. But this can be overcome with checking other linked profile info like blogs attached, because most people who misrepresent themselves may not be consistent in it, and if the data on all online profiles is consistent, it may mean that it's a genuine profile. Another problem is that twitter data cannot represent all the populace.

But from the results of the literature used for this review we can draw the conclusion that the attributes inferred can be applied in a lot of fields. Advertising is one such field, so if a user is marked as a male of age 25 or above, she can be shown more products like aimed at older males. This may prove helpful in providing users with a better experience if demographic specific material can be made available to users easily. User experience can be tailored for individual users based on a mix of latent attributes, e.g., ethnicity, age and gender etc.

Given the present finding, the following are important issues to address in future twitter inference studies. First, researchers can start theorizing on better working machine learning model, so as to improve accuracy and scalability even for the most difficult attribute, like political affiliation. In addition to it, the methodologies used in different researches can be coupled to increase accuracy and infer more attributes apart from age, gender, political affiliation and ethnicity. More research needs to be conducted in incorporating the findings of these and future studies in applications that benefit from user profiling. Also, it must be investigated whether this inference can work for other social media endorsements, for example, the 'like' feature on YouTube or the 'favorite' feature on Instagram and Tumblr. In the future, latent attribute inference can be extended to other social media platforms available online. Another field of research can be inferring political leaning in a multi-party system like India. This can be accomplished using regional political leaning in general, along with data accumulated from political wins in the past.

## Author Profile

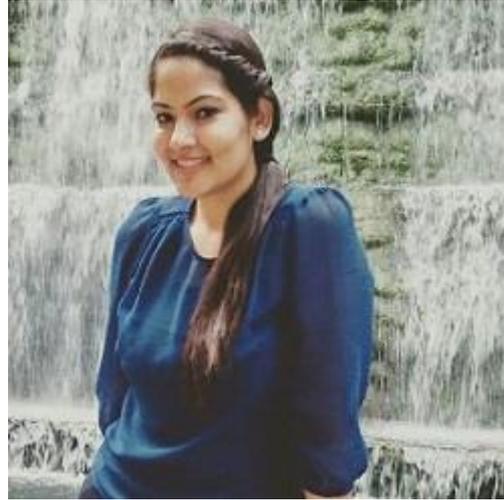

**Surabhi Singh** received the B.Tech. Degree in Computer Science and Engineering from Rajasthan Technical University, Kota in 2014. She has been working as a software engineer with Infosys Limited from August, 2014.